\documentclass[aps,prd,twocolumn,longbibliography,superscriptaddress,floatfix]{revtex4-2}

\usepackage{amsmath,amssymb}
\usepackage{graphicx}
\usepackage{booktabs}
\usepackage{flushend}

\graphicspath{{pic/}}

\begin{document}

\title{Nonperturbative lattice simulation of the nucleation rate in thermal first-order phase transitions}
\author{Jing Zhou}
\affiliation{~Department of Physics and Chongqing Key Laboratory for Strongly Coupled Physics, Chongqing University, Chongqing 401331, P. R. China}

\author{Ligong Bian\thanks{Corresponding Author.}}
\email{lgbycl@cqu.edu.cn}
\affiliation{~Department of Physics and Chongqing Key Laboratory for Strongly Coupled Physics, Chongqing University, Chongqing 401331, P. R. China}

\begin{abstract}
We present a nonperturbative lattice study of vacuum decay through bubble nucleation in a three-dimensional real scalar field theory at finite temperature. The vacuum decay rate is computed by combining an expanded-ensemble multicanonical hybrid Monte Carlo approach, which samples the critical bubble configurations, with real-time Langevin dynamics, which follows the subsequent bubble evolution. We extrapolate the nucleation rate to the continuum limit and compare it with the corresponding one-loop perturbative prediction, finding a discrepancy of about \(6.6\%\).
\end{abstract}

\maketitle

\section{Introduction}
\label{sec:intro}

The first-order phase transition at the electroweak scale is of great interest 
in early-universe cosmology, since if such a transition exists in extensions 
of the Standard Model, it may provide the necessary out-of-equilibrium 
conditions for electroweak baryogenesis~\cite{Kuzmin:1985,Rubakov:1996,Cline:2021,Ellis:2022,Bruggisser:2022,Kanemura:2023,Idegawa:2024},
give rise to the cosmic origin of primordial magnetic fields~\cite{Hogan:1983zz,Di:2020kbw,Enqvist:1993np,Yang:2021uid,Vachaspati:1991nm,Di:2024gsl,Di:2025ncl,Baym:1995fk,Grasso:1997nx,Quashnock:1988vs},
produce primordial black holes~\cite{Liu:2021svg,Franciolini:2025ztf,Carr:2026hot,Lewicki:2023ioy,Gouttenoire:2023naa,Jedamzik:1999am},
and generate primordial gravitational waves observable by space-based 
gravitational-wave detectors~\cite{Hindmarsh:2021,Caprini:2016,Caprini:2019,Auclair:2023}. Moreover, several Pulsar Timing Array collaborations have reported evidence 
for a stochastic gravitational-wave background~\cite{Agazie:2023,Antoniadis:2023,Xu:2023,Reardon:2023,Bian:2020urb,Bian:2023dnv}, 
opening the possibility that new-physics phase transitions contribute to 
the observed signal. The bubble nucleation rate $\Gamma$ is a central input for characterizing the phase transition dynamics. Because the nucleation rate depends exponentially on the height of the 
phase transition barrier, theoretical predictions require the error in 
$|\log\Gamma|$ to be kept under strict control; otherwise, quantitative 
predictions of the gravitational-wave spectrum are not feasible.

Theoretically, the calculation of the nucleation rate originates from the 
saddle-point approximation for zero-temperature quantum tunneling~\cite{Kobzarev:1974,Coleman:1977,Callan:1977} 
and the statistical theory of finite-temperature classical nucleation~\cite{Langer:1967,Langer:1969,Langer:1974,Cahn:1959}. 
In the latter framework, the nucleation rate is factorized into a statistical 
prefactor and a dynamical factor: the former yields the equilibrium 
probability of the critical bubble relative to the metastable vacuum, while 
the latter governs the rate at which the critical bubble traverses the 
barrier top. This formalism was subsequently extended to finite-temperature 
quantum field theory~\cite{Linde:1981,Affleck:1981}. At high temperature, 
the thermodynamics of a four-dimensional quantum field theory reduces to a 
three-dimensional effective field theory~\cite{Kajantie,Braaten}, thereby 
recasting the nucleation rate calculation as a problem in quasi-classical 
statistical field theory.

However, the perturbative expansion exhibits poor convergence at higher 
orders; in particular, the statistical prefactor is dominated by exponentially 
suppressed field configurations, rendering the calculation challenging to 
control. The theoretical uncertainties inherent in perturbative approaches 
have been systematically investigated across a wide range of models~\cite{Kainulainen:2019,Gould:2019,Ekstedt:2020,Croon:2020,Gould:2021b,Ekstedt:2022,Chung:2013,Ekstedt:2022c,Gould:2023,Liu:2026ask,Liu:2025ipj}, 
and the choice of the effective potential and the renormalization 
scale can further introduce additional scheme dependence~\cite{Weinberg:1987,Wainwright:2012,Delaunay:2007,Athron:2022,Zhu:2025pht}. 
These findings underscore the necessity of nonperturbative calculations for 
obtaining a reliable nucleation rate.

Lattice methods provide a systematic nonperturbative approach to the 
statistical prefactor. For weakly metastable systems, direct simulation of 
the thermal nucleation process in real time has been widely 
studied~\cite{Grigoriev:1988,Valls:1990,Alford:1992,Alford:1993,Alford:1993b,Borsanyi:2000,Batini:2024,Pirvu:2023}. An alternative approach is the multicanonical Monte Carlo method~\cite{Berg:1991,Berg:1992}, which enables a fully nonperturbative 
calculation of an extremely small electroweak bubble nucleation 
rate~\cite{Moore:2001}. Nevertheless, the multicanonical method faces a 
sampling obstacle known as the condensation barrier, which is difficult to 
circumvent in large-volume simulations and leads to an exponential decrease 
in simulation efficiency with 
volume~\cite{Binder:1980,Binder:2003}.

Recently, a novel expanded-ensemble method was proposed to effectively 
circumvent the condensation barrier~\cite{Hallfors:2025}. This approach grows 
a bubble through a chain of subvolumes, initiating from a small region where 
the barrier is negligible and progressively enlarging the subvolume until the 
subcritical size is attained, thereby ensuring that the tunneling time no 
longer exhibits exponential growth with volume. Although this method has been 
validated in a two-dimensional scalar field model with a symmetric 
double-well potential, its extension to three dimensions remains unexplored, 
as the form of the potential, the efficiency of the update algorithm, and the 
continuum-limit behavior all warrant further investigation in the 
three-dimensional case.

In this work, we extend the expanded-ensemble multicanonical framework to a 
three-dimensional asymmetric double-well system, enabling the computation of 
the statistical prefactor and the extraction of the critical bubble 
configuration. We employ multicanonical hybrid Monte Carlo (HMC) updates to 
significantly improve sampling efficiency~\cite{Arnold:1999,Duane:1987}. 
Crucially, we introduce a translation-invariant, differentiable weighted 
order parameter $\mathcal{O}_{\rm inv}$ that resolves the fixed-window  pinning issue and obviates the need for auxiliary shift updates~\cite{Hallfors:2025}, 
all while maintaining detailed balance and differentiability for HMC. 
Coupled with real-time Langevin dynamics, this allows us to determine the 
complete nonperturbative nucleation rate and benchmark it against the 
one-loop Langer result. To our knowledge, this constitutes the first 
successful extension of the multicanonical expanded-ensemble method to three 
dimensions.

The structure of this paper is as follows. Section~\ref{sec:method} 
outlines the numerical methodology, while Section~\ref{sec:lattice} details 
the lattice implementation. In Section~\ref{sec:results}, we present the 
complete numerical results and systematically compare the nonperturbative 
nucleation rate with the one-loop perturbative prediction. Finally, 
Section~\ref{sec:conclusion} summarizes our findings and discusses their 
broader implications.

\section{Numerical Method}
\label{sec:method}
In this section, we present the methodology combining the expanded ensemble multicanonical Monte Carlo method with the real-time Langevin technique. The Monte Carlo approach is utilized to extract the statistical prefactor and identify critical bubble configurations. Building upon these configurations, the real-time Langevin dynamics is then applied to evaluate the dynamical prefactor.

\subsection{Order parameter}
\label{sec:order_parameter}

To distinguish between the two phases (symmetric and broken) and identify critical bubbles, an order parameter is needed. The simplest choice is the volume average of the field,
\begin{equation}
\mathcal{O} \equiv \frac{1}{V}\sum_{\vec{x}} \phi(\vec{x}),
\label{eq:mean}
\end{equation}
or its square \(\bar{\phi}^2\), with the latter being more common in current simulation studies. In this paper, we use an improved quadratic-type order parameter,
\begin{equation}
\mathcal{O} = \frac{1}{V} \sum_{\vec{x}} (\phi(\vec{x}) - A)|\phi(\vec{x}) - A|,
\label{eq:linear}
\end{equation}
to be specified below. 
This order parameter not only makes the critical region more prominent, thereby facilitating sampling, but also naturally captures the evolution of the field value, rendering it highly suitable for the HMC updating scheme.

In large-volume simulations, the shift in the order parameter induced by a small bubble can be entirely overwhelmed by bulk fluctuations, rendering the bubble branch in the histogram unresolvable. To address this issue, we introduce a spatial smoothing technique, which replaces the field value with a weighted average of itself and its nearest neighbors, thereby filtering out short-wavelength fluctuations:
\begin{equation}
\phi^{(n+1)}(\vec{x}) = \frac{1}{4}\left( \phi^{(n)}(\vec{x}) + \frac{1}{2}\sum_{\hat{\mu}} \phi^{(n)}(\vec{x}+\hat{\mu}) \right),
\label{eq:smear}
\end{equation}
where \(\hat{\mu}\) runs over positive spatial directions. After \(n\) smearing steps, the smoothed field is denoted by \(\phi_s(\vec{x})\). Smearing strongly narrows the width of the order-parameter distribution in the pure bulk phase, making the bubble signal more visible at large volumes.

\subsection{Condensation barrier}
\label{sec:condensation}

\begin{figure}[!htp]
  \centering
  \includegraphics[width=\columnwidth]{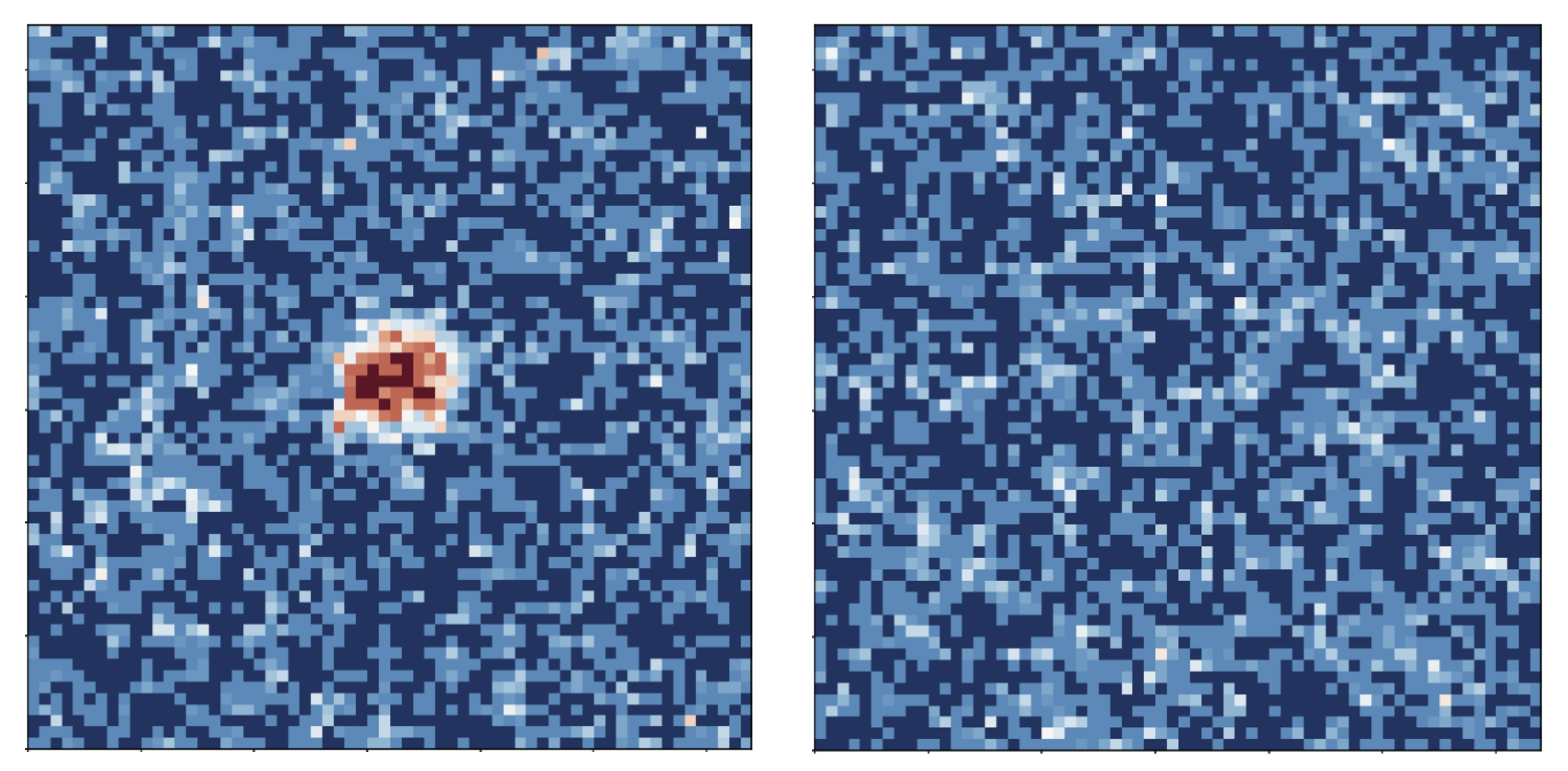}
  \caption{Examples of configurations with approximately the same value of the volume-averaged order parameter. The left panel shows a configuration containing a small bubble (the bubble branch), while the right panel shows a pure bulk configuration (the fluctuation branch).}
  \label{fig:same_order}
\end{figure}

Near the condensation transition point, the same value of the order parameter can correspond to two completely different types of field configurations, referred to as the bubble branch and the fluctuation branch, respectively. 
Fig.~\ref{fig:same_order} illustrates this situation. For the same value of the order parameter, the left panel shows a configuration containing a bubble, whereas the right panel shows a configuration with no bubble, thereby illustrating the distinction between the bubble and fluctuation branches. There is a hidden barrier for transitions between these two branches, as the multicanonical weight depends only on the order parameter and cannot distinguish between them.
As the system volume increases, the physical size of the critical bubble increases, and the height of the condensation barrier increases as well. Periodic boundary conditions can also introduce shape transitions~\cite{Neuhaus:2003}, further increasing the difficulty of sampling.

\subsection{Expanded ensemble method}
\label{sec:expanded}

The general multicanonical framework is reviewed in Appendix~\ref{app:muca}; here we directly present the expanded-ensemble construction used in this work.

The expanded ensemble method \cite{Hallfors:2025} introduces the weighted order parameter and constructs a chain of subensembles, effectively decomposing the rare event in a large volume into a sequence of easily sampled events in small volumes to overcome the condensation barrier.

\subsubsection{Weighted order parameter}

The conventional order parameter \(\mathcal{O}\) is measured over the entire system volume \(V\) (e.g., the full-volume \(\mathcal{O}\)), while the weighted order parameter \(\mathcal{O}^W\) is measured only within a specified subvolume \(V^W \subset V\), e.g., inside a spherical region of radius \(r\). When the subvolume \(V^W\) is small, the condensation barrier for forming a small bubble inside it is also small. Therefore, bubbles can be easily created and destroyed with little condensation barrier, and then the system can accommodate configurations containing larger and larger bubbles by gradually increasing the subvolume.

\subsubsection{Ensemble chain construction}

The entire sampling process is decomposed into a sequence of subensembles \(i = 1, 2, \ldots, N\), each using a different weighted volume \(V_i^W\) and a carefully chosen order-parameter range. The first subensemble uses the smallest subvolume \(V_1^W\), whose order-parameter range extends from the metastable peak \(\mathcal{O}_m\) to the size of the largest bubble that fits inside this small volume. In this small volume, the condensation barrier can be negligible, allowing the field to fluctuate between no-bubble and small-bubble states freely. In the intermediate subensembles, the subvolume \(V_i^W\) is increased step by step, while the order-parameter range of each subensemble is chosen to overlap with those of the neighboring ensembles, ensuring a continuous chain. These subensembles are designed to contain bubbles and prevent their complete evaporation. The final subensemble uses the full volume \(V_N^W = V\), and its order-parameter range is a portion of that of the previous subensemble. Subsequently, the standard multicanonical method can be directly used in the full volume to explore configurations ranging from medium-sized bubbles up to the top of the barrier. Through this chain, a bubble can form in a small window, then grow as the window is enlarged, eventually becoming a stable bubble that does not easily disappear, thus avoiding directly crossing the huge full-volume condensation barrier.

\subsubsection{Translation-invariant differentiable weighted order parameter}
\label{sec:invariant_order_parameter}

The original method employs a weighted order parameter on a fixed subvolume
\begin{equation}
\mathcal{O}_i^W=\frac{1}{V_i^W}\int_{V_i^W} f(\phi)\,dV,
\end{equation}
where $V_i^W$ is a pre-selected spatial region~\cite{Hallfors:2025}. However, the fixed window breaks translational invariance: the bubble tends to be ``pinned'' at the window center, so an additional shift update must be introduced to restore ergodicity of the bubble position. Moreover, the bubble may become stuck at the window edge during simulations, which could cause it to exceed the critical size and lead to simulation failure.
Another approach is to locate the bubble by maximizing, over all possible centers, the convolution of the order parameter field with a weighting function, thereby making the weighted order parameter translation invariant without the shift update. This convolution can be computed efficiently via Fourier transform~\cite{Hallfors:2025}, but the max operation is generally non-differentiable and cannot be used directly in HMC. To this end, we construct a translation-invariant, differentiable weighted order parameter $\mathcal{O}_{\rm inv}$ that fully internalizes the translational degrees of freedom while retaining the differentiability required by HMC.

First, the field is smoothed once; the smoothed field $\phi_s$ preserves spatial features such as the bubble radius and interface position, while suppressing high-frequency fluctuations in the metastable phase. To mark the stable-phase region, we introduce the weight
\begin{equation}
W(\phi_s)=\frac12\left[1+\tanh\left(\alpha(\phi_s-\phi_0^W)\right)\right],
\quad
\phi_0^W=\frac{\phi_{mn}+\phi_{mx}}{2},
\end{equation}
where $O_{mn}$ and $O_{mx}$ are the peak positions of the metastable and stable phases, respectively, and $\alpha$ controls the sharpness of the transition of the weight from the metastable to the stable phase. We take $\alpha=10$, which maintains the distinction between the two phases while avoiding a transition that is too steep and would increase the numerical stiffness of the HMC force. The zero point $\phi_0^W$ is taken at the midpoint of the two phases, so that $W(O_{mn})\approx 0$ and $W(O_{mx})\approx 1$, thereby strongly suppressing the metastable phase in the subsequent centroid encoding. For each direction $k=x,y,z$, we define the complex amplitude
\begin{equation}
Z_k=\sum_x w(\phi_{s,x})e^{i2\pi x_k/L}=X_k+iY_k.
\end{equation}
For a single-bubble configuration, the phase of $Z_k$ directly encodes the bubble centroid position, and the modulus $|Z_k|$ reflects the bubble volume. To avoid numerical singularities when $|Z_k|\to 0$ (no bubble), we introduce the regularization
\begin{equation}
M_k=\sqrt{X_k^2+Y_k^2+\epsilon^2},\qquad
u_k=\frac{X_k}{M_k},\quad v_k=\frac{Y_k}{M_k},
\end{equation}
where $\epsilon^2=1$ is a regularization constant with dimensions of lattice units. The resulting direction vector $(u_k,v_k)$ is a continuous, smooth encoding of the bubble centroid on the periodic torus. Broadcasting the direction vector to each lattice site yields the position field
\begin{equation}
\begin{split}
h(x)&=\sum_k\left[u_k\cos\frac{2\pi x_k}{L}+v_k\sin\frac{2\pi x_k}{L}\right]\\
&=\sum_k\frac{|Z_k|}{M_k}\cos\left(\frac{2\pi x_k}{L}-\theta_k\right),
\end{split}
\end{equation}
where $\theta_k=\arg Z_k$. $h(x)$ represents the proximity of lattice site $x$ to the bubble center. When the whole configuration is translated, $h(x)$ shifts accordingly while maintaining its shape. In the ideal single-bubble case, it approximately satisfies
\begin{equation}
h(x)\simeq 3-\frac{2\pi^2}{L^2}\sum_k(x_k-x_{c,k})^2,
\end{equation}
where $x_c$ denotes the centroid position and $h(x)$ reaches a maximum close to 3 at the centroid and decreases monotonically outward. Then, based on $h(x)$, we construct the soft window
\begin{equation}
g(x)=\frac12\left[1+\tanh\left(\beta(h(x)-c)\right)\right],
\end{equation}
where
\begin{equation}
c=3.0-\frac{2\pi^2\sigma_{\rm window}^2}{L^2},\qquad
\beta=\frac{L^2}{8\pi^2\sigma_{\rm window}}.
\label{eq:soft_window_params}
\end{equation}
$\sigma_{\rm window}$ is the chosen subensemble window radius. Thus, at $r=\sigma_{\rm window}$, $g=1/2$, so the metastable background outside the subvolume is excluded from the order parameter calculation.

Finally, the translation-invariant differentiable weighted order parameter is defined as
\begin{equation}
\mathcal{O}_{\rm inv}(\phi)
=
\frac{
\sum_x g(x;\phi_s)\,f(\phi_{s,x})
+
\epsilon_{\rm reg}\,
\frac{1}{V}\sum_x f(\phi_{s,x})
}{
\sum_x g(x;\phi_s)+\epsilon_{\rm reg}
},
\label{eq:Oinv}
\end{equation}
where $V=L^3$, $f(\phi_s)=(\phi_s-\phi_0^f)|\phi_s-\phi_0^f|$ is the local contribution function, and $\phi_0^f=O_{mn}$ makes the zero point of the order parameter correspond to the pure metastable background. The regularization term $\epsilon_{\rm reg}=10^{-3}$ is a dimensionless numerical constant, used to prevent the denominator from becoming too small when the window is completely closed; when the window is completely closed, $\mathcal{O}_{\rm inv}$ reduces to the global average $\frac{1}{V}\sum_x f(\phi_{s,x})$.

Under periodic boundary conditions, the window center automatically follows the bubble, so $\mathcal{O}_{\rm inv}$ remains unchanged when the bubble drifts freely in the system. It can also be verified that $\mathcal{O}_{\rm inv}\in C^1(\phi)$, that its gradient $\nabla_\phi\mathcal{O}_{\rm inv}$ is continuous, and that it is accurately computed via adjoint propagation. The forward smoothing and the adjoint kernel strictly match, the forces can be obtained analytically, and HMC detailed balance is strictly preserved. Therefore, this order parameter is naturally suited for multicanonical HMC (MCHMC) updates in the expanded ensemble method. It avoids both the fixed-window limitation of the original method and the non-differentiability of the Fourier transform maximum scheme, providing a simpler and more efficient route for expanded ensembles within the HMC framework.

\subsubsection{Update algorithms}

\begin{figure}[htbp]
  \centering
  \includegraphics[width=\columnwidth]{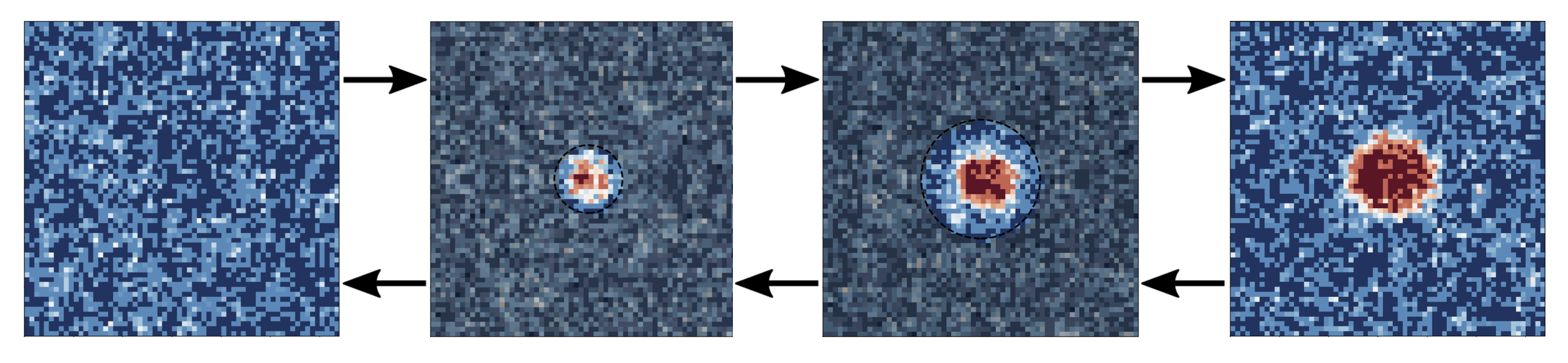}
  \caption{Illustration of the ensemble swap update between adjacent subensembles.}
  \label{fig:exchange}
\end{figure}

To move the system along the chain, the algorithm combines two types of updates. Within each subensemble \(i\), standard multicanonical updates are used to change the field configuration and explore the allowed order-parameter range; the weight \(W_i\) is designed to flatten the histogram of the weighted order parameter \(\mathcal{O}_i^W\), which denotes \(\mathcal{O}_{\rm inv}\) within subensemble \(i\). In addition, ensemble swap updates attempt to swap between adjacent subensembles \(i\) and \(j\). The acceptance probability for such a swap is
\begin{equation}
\begin{split}
P(i \rightarrow j) 
&= \min\Big(1, \exp\Big[-\big(W_j(\mathcal{O}_j^W) - W_i(\mathcal{O}_i^W)\big) \\
&\qquad - (\eta_j - \eta_i)\Big]\Big),
\end{split}
\label{eq:swap}
\end{equation}
where \(\eta_i\) are bias parameters adjusted in a later iteration to balance the visitation probabilities of the different subensembles. When the configuration lies at the boundary of the overlap region of the two ensembles, the acceptance probability is close to unity. A schematic illustration of the ensemble swap update is shown in Fig.~\ref{fig:exchange}.

\subsubsection{Iteration procedure}

The iteration employs a two-step iterative procedure. First, the multicanonical weights \(W_i\) are iterated independently for each subensemble \(i\) to obtain the respective weight functions to ensure each subensemble uniformly visits its allowed order-parameter range. After fixing all \(W_i\), the relative weights \(\eta_i\) are iteratively adjusted so that the system visits each subensemble index \(i\) with equal probability, ensuring that all subensembles on the chain are visited equally, thereby obtaining global statistics that are as accurate as possible.

\subsubsection{Recovering the physical probability}

After the above procedure, one obtains the multicanonical probability distribution \(P_{\text{muc}}\). By reweighting, the canonical (physical) probability distribution \(P_{\text{can}}\) is obtained.
In the simulation, the condensation barrier only appears in the first small-volume subensemble, where it is negligible. Therefore, the total simulation time does not grow exponentially with the volume, and the statistical part of the nucleation rate is
\begin{equation}
 P_c^{normalised} 
=
\frac{P\big(|\mathcal{O}-\mathcal{O}_{C}|<\epsilon/2\big)}
{\epsilon\, P\big(\mathcal{O}<\mathcal{O}_{C}\big)}.
\label{eq:Pc_normalised}
\end{equation}

\subsection{Real-time Langevin dynamics and complete nucleation rate}
\label{sec:langevin}

The statistical weight gives the exponential suppression factor, but to obtain the complete nucleation rate, one must compute the dynamical factor near the critical point \cite{Langer:1969,Moore:2001}. Based on the critical field configurations collected from the multicanonical simulation, the real-time dynamics is governed by the Langevin equations in Eq.~\eqref{eq:langevin}, with the effective Hamiltonian
\begin{equation}
H_{\mathrm{eff}}
=
\int d^3x
\left[
\frac12 \pi^2
+\frac12 (\partial_i \phi)^2
+V_3(\phi)
\right].
\end{equation}

For each critical configuration, we compute the flux across the barrier,
\begin{equation}
\langle \mathrm{flux} \rangle
= \left\langle \left| \frac{\Delta \mathcal{O}}{\Delta t} \right|_{\mathcal{O}_c} \right\rangle,
\end{equation}
which represents the average rate of change of the order parameter near the critical value \(\mathcal{O}_c\). We employ the smoothed nonlinear order parameter
\begin{equation}
\mathcal{O}[\phi]
=
\frac{1}{V}\sum_x
\bigl(\phi_s(x)-A\bigr)
\bigl|\phi_s(x)-A\bigr|,
\qquad
\phi_s=S\phi,
\label{eq:smooth_nonlinear_op}
\end{equation}
where \(S\) is the separable binomial smoothing operator and \(A\) is the peak position of the metastable-phase histogram. Since the momentum field is Gaussian, \(\langle \pi(x)\pi(y)\rangle=\delta_{xy}\), the conditional distribution of \(\dot{\mathcal{O}}\) at fixed \(\phi\) is Gaussian with zero mean and variance
\begin{equation}
\sigma^2
=
\frac{1}{V^2}
\sum_x
\Bigl[
S f'\bigl(\phi_s-A\bigr)
\Bigr]_x^2,
\end{equation}
where \(f(u)=u|u|\), so that \(f'(u)=2|u|\). Here the subscript \(_x\) denotes the value of the field obtained after applying \(S\). The flux through the critical surface \(\mathcal{O}=\mathcal{O}_c\) is therefore
\begin{equation}
\left\langle
\left|\frac{\Delta \mathcal{O}}{\Delta t}\right|
\right\rangle_{\mathcal{O}_c}
=
\sqrt{\frac{8}{\pi V^2}}
\left\langle
\sqrt{
\sum_x
\Bigl[
S\bigl(|\phi_s-A|\bigr)
\Bigr]_x^2
}
\right\rangle_{\mathcal{O}_c}
\label{eq:flux_general}
\end{equation}
where \(\langle\cdots\rangle_{\mathcal{O}_c}\) denotes the average over configurations constrained to \(\mathcal{O}=\mathcal{O}_c\). In the unsmoothed case \(S=I\), Eq.~\eqref{eq:flux_general} reduces to
\begin{equation}
\left\langle
\left|\frac{\Delta \mathcal{O}}{\Delta t}\right|
\right\rangle_{\mathcal{O}_c}
=
\sqrt{\frac{8}{\pi V}}
\left\langle
\sqrt{
\frac{1}{V}\sum_x
\bigl(\phi(x)-A\bigr)^2
}
\right\rangle_{\mathcal{O}_c}.
\label{eq:flux_unsmoothed}
\end{equation}
For the quadratic order parameter \(\phi^2_{\rm av}\), taking \(A=0\) and positive fields, one recovers the familiar result~\cite{Moore:2001b}
\begin{equation}
\langle |\dot{\phi^2_{\rm av}}|\rangle_{\phi^2_{{\rm av},c}}
=
\sqrt{\frac{8\phi^2_{{\rm av},c}}{\pi V}}.
\end{equation}

In addition, each configuration is evolved both forward and backward in time, with the initial momenta reversed for the backward evolution. We count the fraction of trajectories that truly cross the barrier by defining
\begin{equation}
{d} = \frac{\delta_{\mathrm{tunnel}}}{N_{\mathrm{crossings}}},
\end{equation}
where \(\delta_{\mathrm{tunnel}} = 1\) if the trajectory starts and ends in different phases and \(0\) otherwise, and \(N_{\mathrm{crossings}}\) is the number of times the trajectory crosses \(\mathcal{O}_c\). A factor of \(1/2\) is included to count only metastable \(\to\) stable trajectories, avoiding double counting.
Although \(\langle |\Delta \mathcal{O}/\Delta t| \rangle\) and \(d\) individually depend on the Langevin time step, their product converges in the short-time limit.

From the Langevin simulations of critical configurations, we measure the velocity factor \( \langle \mathrm{flux} \rangle \) and the transmission coefficient \(d\). The product of these two quantities gives the dynamical prefactor. Although the individual factors vary when the step is halved, their product remains constant within errors, as expected from the factorization of the nucleation rate in Langer's theory~\cite{Langer:1969}. The dynamical factor \(A_{\mathrm{dyn}}\) is decomposed into two physically distinct quantities:
\begin{equation}
A_{\mathrm{dyn}} \leftrightarrow
\langle \mathrm{flux} \rangle
\times d.
\end{equation}
Therefore, the final form of the nucleation rate in the numerical simulation is~\cite{Moore:2001b,GKW}:
\begin{equation}
{\Gamma}
=
\frac{1}{2V}\,
 P_c^{normalised} 
\,
\left\langle \left|\frac{\Delta \mathcal{O}}{\Delta t}\right|\right\rangle
\,
d.
\label{eq:rate_full}
\end{equation}

The flux-over-population construction ensures that the final rate is independent of the order parameter, as long as the dividing surface separates the two phases~\cite{Moore:2001}. We therefore regard \(P_c^{\mathrm{normalised}}\), \(\langle|\Delta\mathcal{O}/\Delta t|\rangle\), and \(d\) as coordinate-dependent intermediate quantities and do not attach separate physical meaning to them.

\section{Lattice details}
\label{sec:lattice}
In this section, we present the model for lattice simulation and the simulation setup.
\subsection{Lattice action and parameters}
\label{sec:lattice_action}

\begin{figure}[!htp]
  \centering
  \includegraphics[width=0.8\linewidth]{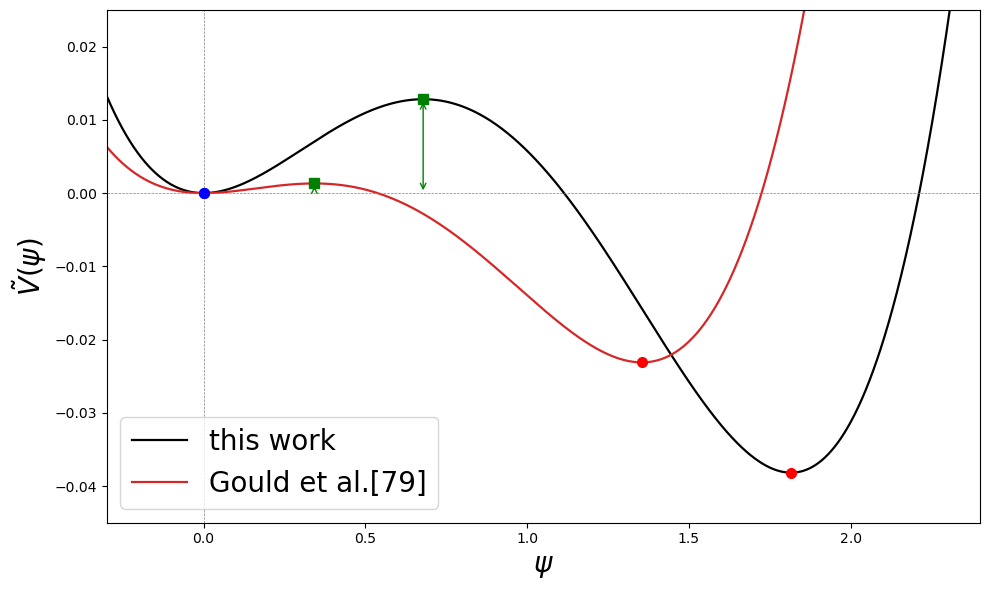}
  \caption{This shows the difference between the potential used in this work and that in Ref.~\cite{GKW}. The blue, green, and red dots indicate the false vacuum, the barrier, and the true vacuum, respectively.}
  \label{fig:potential}
\end{figure}

To investigate the bubble nucleation rate for the first-order phase transition in this work, the lattice 3D action takes the form
\begin{equation}
S_3[\phi] = \int d^3x \left[ \frac12 (\partial_i \phi)^2 + V_3(\phi) \right],
\end{equation}
with the 3D effective potential~\cite{Gould:2021}:
\begin{equation}
\begin{split}
V_3(\phi)  
&= \frac12 m_3^2 \phi^2 + \frac16 g_3 \phi^3 + \frac{1}{24} \lambda_3 \phi^4, 
\label{eq:V3_main}
    \end{split}
\end{equation}
where 
\begin{equation}
m_3^2 = \gamma (T_3^2 - (T_3^0)^2),  g_3 = - 2 \alpha T_3^{3/2}, \lambda_3 = 6 \lambda T_3.
\end{equation}
The 3D dimensionless parameters used in this work are
\begin{equation}
\begin{split}
\gamma =& 1.17,
\quad \alpha T_3 = 1.45, 
\quad \lambda T_3^{1/2} = 0.5, \\
&\quad T_3^0 = 0.80, 
\quad T_3 = 1.49\;.
\end{split}
\end{equation}
Fig.~\ref{fig:potential} compares the potential used in this work with that of Ref.~\cite{GKW}. Although the two potentials are not normalised to the same dimensionless scale, the barrier obtained here is higher, indicating that nucleation is more strongly suppressed.

Considering the influence of lattice discretisation, the lattice action takes the form
\begin{equation}
\begin{split}
S_{\rm lat}
&=\sum_x\Big[
-\frac12 Z_\phi\,\phi_x(\nabla^2_{\rm lat}\phi)_x
+\sigma_{\rm lat}\phi_x \\
&\qquad+\frac12 Z_\phi Z_m\,m^2_{\rm lat}\phi_x^2
+\frac{1}{4!}Z_\phi^2\lambda_{\rm lat}\phi_x^4
\Big],
\end{split}
\label{eq:lattice_action}
\end{equation}
where \(Z_\phi\) and \(Z_m\) are multiplicative renormalisation constants, and \(\sigma_{\rm lat}\) is the coefficient of the linear term after the shift. The counterterms \(\delta\lambda\) and \(\delta m^2\) are included in the definitions of \(\lambda_{\rm lat}\) and \(m^2_{\rm lat}\). The derivation of the dimensionally reduced action, the field shift, the matching relations, and the overcooling procedure is presented in Appendix~\ref{app:lattice_matching}. 

\subsection{Simulation setup}
\label{sec:simulation_setup}

Here we show the lattice geometry, the spacing choices, and the sampling parameters of both the expanded-ensemble MCHMC simulation and the real-time Langevin simulation. 

\subsubsection{Lattice geometry and spacing}

All simulations use periodic boundary conditions on a cubic lattice with \(N_s^3\) sites and physical volume \(L^3\). We keep the dimensionless physical volume approximately fixed,
$L\lambda_3 \sim 128$,
so that the lattice spacing \(a\lambda_3\) and the number of sites are related as
\begin{equation}
N_s = \frac{L}{a} = \frac{L\lambda_3}{a\lambda_3}.
\end{equation}
The spacings used in this work are \(a\lambda_3 \in \{1.5,\,2.0,\,2.5,\,3.0\}\), which correspond to \(N_s = 86,\ 64,\ 52,\ 42\) respectively.  

The range of \(a\lambda_3\) is chosen so that the \(O(a^3)\) continuum extrapolation is reliable. The smallest spacing is limited by the cost of increasing \(N_s=L/a\) and also by the validity of the high-temperature dimensional reduction, which requires
\begin{equation}
a \gtrsim \frac{1}{2\pi T} \sim \frac{1}{T}.
\label{eq:amin_T}
\end{equation}
 For our parameters \(T_3=\text{1.49}\), this bound corresponds to \(a\lambda_3 \gtrsim \text{1.0}\). 

\begin{table}[!htp]
\centering
\caption{Statistical lattice simulation parameters for each spacing.}
\begin{tabular}{cccccc}
\toprule
\(a\lambda_3\) & \(r_{ini}/a\) & \(\alpha\) & \(N_{\rm ens}\) & \( \Delta t\)& \(N_{\rm MD}\) \\
\midrule
1.5& 6& 1.15& 9& 0.12& 15\\
2.0& 6& 1.15& 7& 0.12& 10\\
2.5& 5& 1.15& 3& 0.10& 8\\
3.0& 5& 1.10& 6& 0.10& 5\\
\bottomrule
\end{tabular}
\label{tab:stat_params}
\end{table}

\begin{figure*}[!htp]
  \centering
  \includegraphics[width=0.75\linewidth]{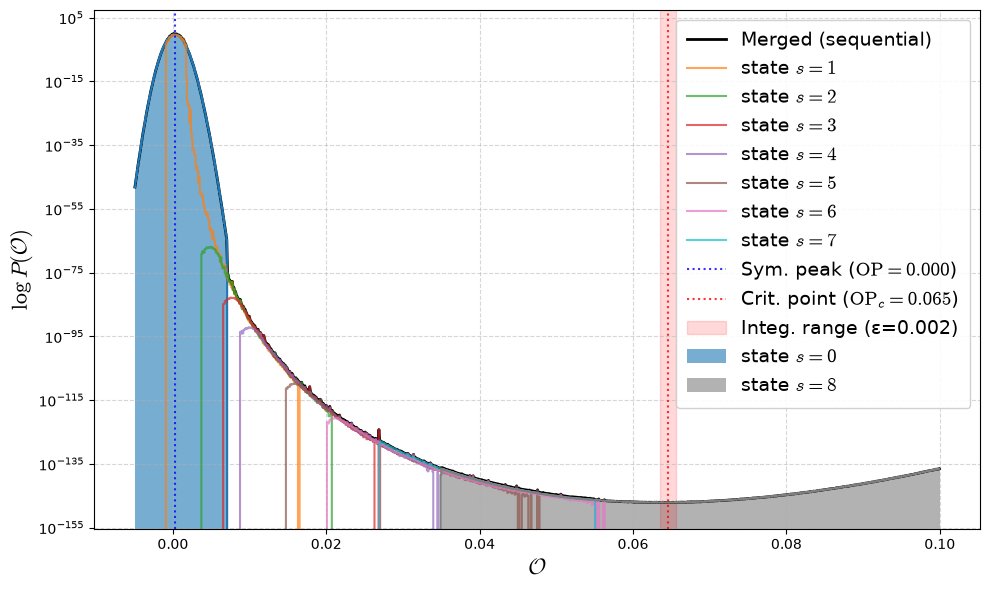}
  \caption{Reweighted physical probability distribution \(P(\mathcal{O})\) obtained from the expanded-ensemble multicanonical simulation with \(a \lambda_3 = 3\). The different curves correspond to individual subensembles, which are combined after normalisation.}
  \label{fig:result}
\end{figure*}

\subsubsection{Expanded-ensemble MCHMC parameters}

The expanded-ensemble chain consists of \(N_{\rm ens}\) subensembles, with both \(N_{\rm ens}\) and the subvolume radii varying with the lattice spacing. The first subensemble uses the smallest subvolume, taken to be a sphere of radius \(r_{ini}\), and subsequent subensembles enlarge the subvolume by a fixed growth ratio \(\alpha = {r_{i+1}}/{r_i}\), so that the \(i\)-th subensemble uses a sphere of radius \(r_i = r_{ini} (\alpha)^{i-1}\), increasing until the subcritical size is reached.
The molecular-dynamics parameters also depend on the lattice spacing and are tuned to balance simulation efficiency against maintaining an average HMC acceptance rate of about \(\text{60}\%\). Here \(\Delta t\) is the HMC molecular-dynamics time step and \(N_{\mathrm{MD}}\) is the number of molecular-dynamics steps per trajectory. The corresponding per-spacing parameters are collected in Table~\ref{tab:stat_params}. The production runs contain around \(N_{\rm conf} \sim 10 ^6\) configurations per subensemble. Statistical errors are estimated by jackknife resampling over contiguous time blocks of the expanded-ensemble chain: the chain is divided into \(20\) blocks, and each jackknife sample removes the same time block from all subensembles before recomputing \(P_c^{\mathrm{normalised}}
\).

\subsubsection{Real-time Langevin parameters}

The dynamical prefactor is extracted from real-time Langevin evolution of critical configurations. For each lattice spacing, we use \(N_{\rm crit} \sim 150\) critical configurations selected from the multicanonical simulation in the interval \(|\mathcal{O}-\mathcal{O}_{C}|<\epsilon/2\). The Langevin equations are integrated with time step \(\Delta t_{\rm Lang} = \text{0.05}\), using a damping coefficient \( \gamma = 1/L\)~\cite{Moore:2001b}. For each configuration, we perform 30 independent Langevin runs, each evolved forward and then backward, with the momenta reversed at the start of the backward evolution. The statistical error on the \( \langle \mathrm{flux} \rangle \) is estimated from the config-to-config scatter over the  \(N_{\rm crit} \) configurations, while that on \(d\) is estimated from a Poisson bootstrap over the YES/NO outcomes.

\section{Results}
\label{sec:results}

In this section we present the numerical results for the statistical factor, the dynamical factor, and the final nucleation rate. The statistical factor is obtained from the expanded-ensemble MCHMC simulation and extrapolated to the continuum limit, while the dynamical factor is extracted from real-time Langevin evolution of critical configurations.

\subsection{Statistical factor}
\label{sec:stat_factor}

The reweighted physical probability distribution \(P(\mathcal{O})\) at \(a \lambda_3 = 3\) is shown in Fig.~\ref{fig:result}. The different curves correspond to the individual subensembles of the expanded-ensemble chain, which overlap in the order-parameter range and are combined into a single continuous distribution after normalisation. The left peak corresponds to the symmetric phase, while the intermediate region around the local minimum of \(P\) is the droplet branch from which the critical bubble configuration is extracted. The normalised critical-bubble probability \(P_c^{\mathrm{normalised}}\) is obtained from Eq.~\eqref{eq:Pc_normalised}.

\begin{figure}[!htp]
  \centering
  \includegraphics[width=0.8\columnwidth]{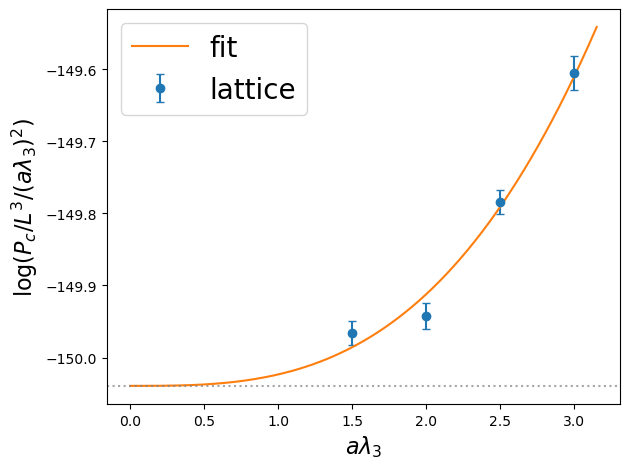}
  \caption{The zero-lattice-spacing extrapolation with cubic fits.}
  \label{fig:fit}
\end{figure}

Since the lattice action is \(O(a^4)\)-improved, the leading lattice-spacing dependence of the logarithm of the rate arises at \(O(a^3)\). A least-squares fit of the form
\begin{equation}
\log P_c^{\mathrm{normalised}}
=
f_0 + c\,(a\lambda_3)^3
\label{eq:fit_stat}
\end{equation}
to the four data points yields, in the continuum limit,
\begin{equation}
\log P_c^{\mathrm{normalised}}
=
-150.039 \pm 0.025 ,
\label{eq:Pc_cont}
\end{equation}
where the error has been inflated by \(\sqrt{\chi^2/\mathrm{dof}}= 1.498\)  to account for the fit quality. The best-fit line and the lattice data are shown in Fig.~\ref{fig:fit}.

\subsection{Dynamical factor}
\label{sec:dyn_factor}

The dynamical factor \(A_{\mathrm{dyn}}\) is extracted from real-time Langevin evolution of critical configurations, as described in Sec.~\ref{sec:langevin}. For each lattice spacing, we select \(N_{\mathrm{crit}}\sim 150\) configurations in the interval \(|\mathcal{O}-\mathcal{O}_C|<\epsilon/2\) and evolve each configuration both forward and backward in time, with initial momenta reversed for the backward evolution. From these trajectories we determine the flux \(\langle|\Delta\mathcal{O}/\Delta t|\rangle\) and the effective tunnelling fraction \(d\).

\begin{figure}[htbp]
  \centering
  \includegraphics[width=0.8\columnwidth]{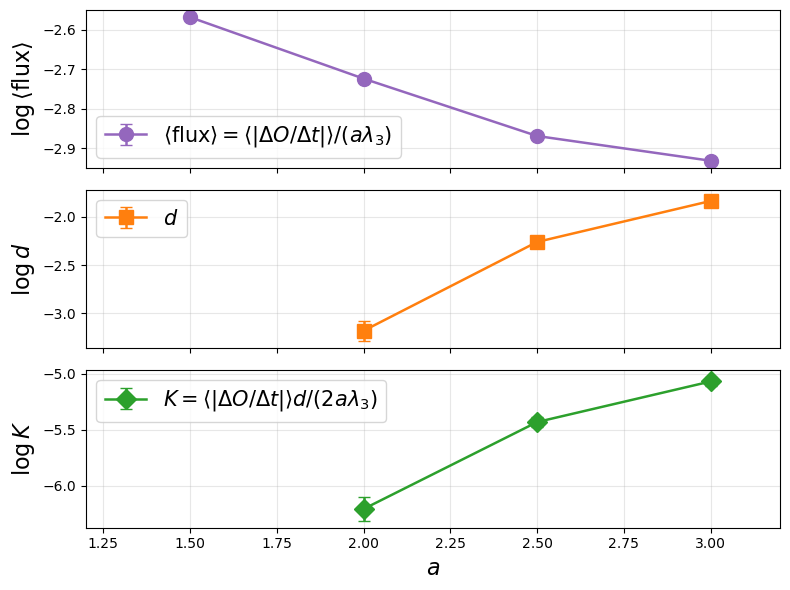}
  \caption{\( \langle \mathrm{flux} \rangle \), \(d\) and dynamical factors at different lattice spacings. The upper panel shows the  \( \langle \mathrm{flux} \rangle \) , the middle panel shows \(d\), and the lower panel shows the full dynamical factor.}
  \label{fig:D}
\end{figure}

Fig.~\ref{fig:D} shows the resulting dynamical factor at different lattice spacings. The flux depends smoothly on the lattice spacing, increasing slightly as the lattice spacing is reduced. The tunnelling fraction \(d\), on the other hand, decreases significantly as the lattice spacing is reduced. At the smallest spacing, \(a\lambda_3=1.5\), we evolved 800 critical configurations, producing 24\,000 real-time trajectories in total, none of which successfully complete a full metastable-to-stable crossing; instead, they eventually evolve into the true vacuum. We conjecture the reason is that \(d\) at \(a\lambda_3=1.5\) is so low that it cannot be detected in this simulation. The factor $d$ has been taken to be a constant in Ref.~\cite{Moore:2001b}.

\subsection{Final nucleation rate}
\label{sec:rate_combined}

\begin{figure}[!htp]
  \centering
  \includegraphics[width=0.8\columnwidth]{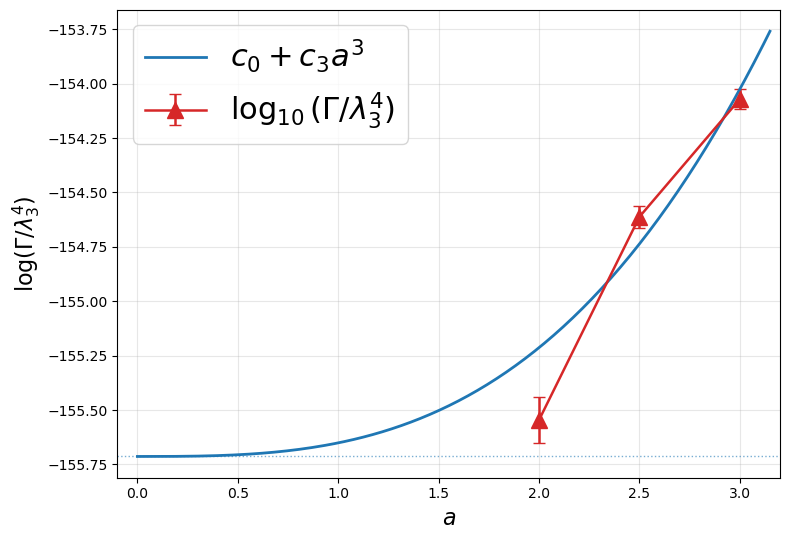}
  \caption{The zero-lattice-spacing extrapolation of the final nucleation rate.}
  \label{fig:final}
\end{figure}

Combining the statistical and dynamical factors yields the final nucleation rate, shown in Fig.~\ref{fig:final}. Because of the limited dynamical factor data, we can fit only three points, and the fit quality is poor as \(\sqrt{\chi^2/\mathrm{dof}}= 4.161\); nevertheless, it can still be seen that the nucleation rate decreases as \(a \lambda_3\) decreases, which is opposite to the trend in Ref.~\cite{GKW}. The continuum-extrapolated nucleation rate per unit volume is
\begin{equation}
\log \Gamma^{\mathrm{lat}}
=
-155.713 \pm 0.430.
\label{eq:rate_combined}
\end{equation}

Using the numerical package {\it BubbleDet}~\cite{Ekstedt:2023}, we compute the full one-loop theoretical result:
\begin{align}
e^{-\Delta S[\phi_S]}&\simeq 9.989\times 10^{-113} ,
A_{\mathrm{pre}}\simeq 2.000\times 10^{-33},\\
A_{\mathrm{stat}}
&=
A_{\mathrm{pre}} 
e^{-\Delta S[\phi_S]}
\simeq
1.997\times 10^{-145},\\
A_{\mathrm{dyn}}& \simeq 4.524\times 10^{-2} \;.
\label{eq:Astat_1loop}
\end{align}
The corresponding theoretical nucleation rate is
\begin{equation}
\Gamma^{\mathrm{1\text{-}loop}}
=A_{\mathrm{stat}} \times A_{\mathrm{dyn}}
=
9.036\times 10^{-147} .
\label{eq:rate_1loop}
\end{equation}
See Appendix~\ref{app:comparison} for details. Hence the difference in \(|\log\Gamma|\) is 
\begin{equation}
\frac{\bigl|\log\Gamma^{\mathrm{lat}}-\log\Gamma^{\mathrm{1\text{-}loop}}\bigr|}{\bigl|\log\Gamma^{\mathrm{1\text{-}loop}}\bigr|}
=
6.6\%,
\label{eq:diff_logGamma}
\end{equation}
This is consistent with the finding of Ref.~\cite{GKW,GKWErratum} that one-loop perturbation theory and lattice simulations differ by about \(9\%\) in \(|\log\Gamma|\) in a real scalar theory with a tree-level barrier, even in a parameter region where perturbation theory is expected to perform well. Part of this difference may be due to the one-loop approximation, and a detailed comparison is left for future work.

With the physical volume $L\lambda_3 \simeq 128$ fixed across all four spacings, finite-volume corrections are common to all data points and cannot bias the functional form of the continuum extrapolation. Moreover, the critical bubble radii $r_c$ in this work satisfy $r_c/L \simeq 1/4$, corresponding to a volume fraction below the $15\%$ threshold for periodic-image effects~\cite{Moore:2001b}, and the volume dependence of the nucleation rate is exponentially suppressed rather than following a power law~\cite{GKW}. Therefore, we regard the residual finite-volume correction as negligible relative to the statistical error on $\log P_c^{\mathrm{normalised}}$.

\section{Conclusion and Discussion}
\label{sec:conclusion}

We present a fully nonperturbative calculation of the bubble nucleation rate for a first-order phase transition in a three-dimensional scalar field theory with an asymmetric double-well potential. In doing so, we extend the expanded-ensemble multicanonical method to three dimensions for the first time. The statistical prefactor is obtained from HMC updates~\cite{Arnold:1999}, while the dynamical prefactor is extracted from real-time Langevin evolution. Combining the two yields the full nucleation rate, which we compare with the perturbative result computed with BubbleDet~\cite{Ekstedt:2023}.

Although we retain the standard decomposition of the nucleation rate into a statistical and a dynamical factor~\cite{Moore:2001,Moore:2001b,Gould:2022}, our method overcomes the condensation barrier and the sampling bottleneck that plague standard multicanonical simulations at large volumes by employing an expanded-ensemble MCHMC with a translation-invariant, differentiable weighted order parameter. Within the same three-dimensional real scalar framework as in Ref.~\cite{GKW}, we additionally consider a different potential and a different choice of parameters, thereby providing a nonperturbative benchmark that is independent of the perturbative calculation and allowing the statistical and dynamical contributions to be separated systematically. The resulting nonperturbative rate, $ \Gamma \simeq 1.936\times10^{-156} $ in lattice units, differs from the BubbleDet result by $6.6\%$ in $ |\log\Gamma| $. This is close to the difference of about $9\%$ in $ |\log\Gamma| $ reported in Ref.~\cite{GKW,GKWErratum} for the parameter choices considered there. Such comparisons are useful for testing the convergence of perturbation theory and provide a well-defined target for future two-loop calculations.

Physically, the nucleation rate obtained in this work provides a
nonperturbative input for predictions of the gravitational-wave signal from
first-order electroweak phase transitions, helping to reduce theoretical
uncertainties relevant to LISA and pulsar-timing-array
observations~\cite{Caprini:2016,Caprini:2019,Auclair:2023,Agazie:2023,Antoniadis:2023,Xu:2023}.
It also governs the degree of supercooling and the duration of the transition,
and therefore influences the efficiency of electroweak
baryogenesis~\cite{Cline:2021,Ellis:2022,Bruggisser:2022}. Methodologically,
the expanded-ensemble HMC framework can in principle be extended to other rare
processes characterized by a condensation barrier and a suitable
differentiable order parameter, such as topological-defect formation and
nucleation in more complex gauge--Higgs systems.

The present study is restricted to scalar fields and does not include gauge
fields or fermions. A more systematic investigation of the continuum limit and
of finite-volume effects~\cite{Neuhaus:2003} is likewise desirable. Future
applications of the method include its extension to more realistic
electroweak models, comparison of the results with two-loop perturbative
calculations, and the use of the nonperturbative nucleation rate as an input
for gravitational-wave spectrum simulations, enabling further cosmological
predictions and tests.

\begin{acknowledgments}

This work was supported in part by the National Natural Science Foundation of China under Grant Nos. 12675130, 12322505, 12547101, and Fundamental Research Funds for the Central Universities Grant No. 2026CDJZKPT-10.
L.B. also acknowledges Chongqing Natural Science Foundation under Grant No. CSTB2024NSCQ-JQX0022.  
We also acknowledge the use of the HILA lattice framework for the numerical simulations presented in this work and thank its developers for making it publicly available.

\end{acknowledgments}

\bibliography{references}

\appendix

\section{Review of Langer theory}
\label{app:comparison}

For a thermal first-order phase transition, the bubble nucleation rate can be described by Langer's finite-temperature field theory \cite{Langer:1967,Langer:1969,Langer:1974,Cahn:1959}. The long-wavelength dynamics of the scalar field is governed by the Langevin equations
\begin{equation}
\partial_t \phi = \pi, \qquad
\partial_t \pi = -\frac{\delta H_{\mathrm{eff}}}{\delta \phi} - \gamma \pi + \xi,
\label{eq:langevin}
\end{equation}
where \(\phi\) is the 3D effective scalar field, \(\pi\) its canonical momentum, \(H_{\mathrm{eff}}\) the effective Hamiltonian constructed to match the long-wavelength equal-time correlation functions of the full quantum theory\cite{Berera:2019,Gould:2021}, \(\gamma\) the damping rate, and \(\xi\) a Gaussian noise satisfying the fluctuation-dissipation relation
\begin{equation}
\langle \xi(t,\mathbf{x})\xi(t',\mathbf{x}')\rangle
= 2\gamma \delta(t-t')\delta^{(3)}(\mathbf{x}-\mathbf{x}').
\end{equation}
The probability density \(P(\phi,\pi)\) obeys a Fokker-Planck equation \cite{Berera:2019}. Assuming the metastable phase is sufficiently long-lived, the nucleation rate per unit volume is given by the steady probability flux over the barrier \cite{Langer:1969,Kramers:1940} and factorizes as
\begin{equation}
\Gamma = A_{\mathrm{dyn}} \times A_{\mathrm{stat}},
\label{eq:factorization}
\end{equation}
where \(A_{\mathrm{stat}}\) is the statistical factor, corresponding to the determinant of fluctuations around the critical bubble, and \(A_{\mathrm{dyn}}\) is the dynamical factor, relating to the flux at the top of the barrier and the transmission probability \cite{Gould:2021,Gould:2022,Berera:2019}.

The statistical factor is obtained from a saddle-point expansion about the critical bubble \(\phi_S\) \cite{Coleman:1977,Callan:1977,Linde:1981,Affleck:1981},
\begin{equation}
A_{\mathrm{stat}} = A_{\mathrm{pre}} \, e^{-\Delta S[\phi_S]},
\label{eq:Astat}
\end{equation}
where the fluctuation prefactor is
\begin{equation}
A_{\mathrm{pre}} =
\left(\frac{\Delta S[\phi_S]}{2\pi}\right)^{3/2}
\sqrt{\left|
\frac{\det(S''[\phi_0]/2\pi)}
{\det'(S''[\phi_S]/2\pi)}
\right|},
\label{eq:Atilde}
\end{equation}
where \(\phi_0\) is the metastable phase, \(S''\) denotes the second functional derivative of the action, the prime on \(\det'\) means that zero modes are omitted \cite{Ekstedt:2023}, and
\begin{equation}
\Delta S[\phi_S] \equiv S[\phi_S] - S[\phi_0].
\end{equation}
The dynamical factor is
\begin{equation}
A_{\mathrm{dyn}} =
\frac{1}{2\pi}
\left(
\sqrt{|\lambda_-| + \frac{\gamma^2}{4}}
- \frac{\gamma}{2}
\right),
\label{eq:Adyn}
\end{equation}
where \(\lambda_-\) is the single negative eigenvalue of \(S''[\phi_S]\) and \(\gamma\) is the damping rate \cite{Langer:1974,Gould:2021,Gould:2022}. These expressions are the one-loop perturbative results; in our lattice calculation we extract the nonperturbative analogues directly \cite{Moore:2001,Moore:2001b,Gould:2022}.

\section{Free energy}
\label{app:free_energy}

The reweighted probability distribution \(P(\mathcal{O})\) is shown in Fig.~\ref{fig:result}. The free energy is defined as
\begin{equation}
F(\mathcal{O}) = -T \ln P(\mathcal{O}),
\end{equation}
where \(\mathcal{O}\) is an order parameter that measures the degree of the phase transition and distinguishes the symmetric and broken phases. In the dimensionally reduced 3D effective theory, the partition function is
\begin{equation}
Z = \int \mathcal{D}\phi\, e^{-S_3[\phi]},
\end{equation}
with \(S_3\) the dimensionless 3D effective action. The constrained free energy is then \cite{Berg:1991,Berg:1992}
\begin{equation}
F(\mathcal{O}_{0})
=
-T \ln \int \mathcal{D}\phi\, e^{-S_3[\phi]}\,
\delta(\mathcal{O}-\mathcal{O}_{0}).
\end{equation}
The difference between the free energy at the separatrix and that of the metastable minimum defines the dimensionless suppression exponent
\begin{equation}
S \equiv \frac{F_C - F_{\text{symm}}}{T},
\label{eq:S_value}
\end{equation}
where \(F_C\) is the free energy at the separatrix, corresponding to the critical bubble configuration, and \(F_{\text{symm}}\) is the free energy of the metastable phase.

In the saddle-point approximation, the free-energy difference is given by the classical action difference,
\begin{equation}
F_C - F_{\text{symm}} \approx T\,\Delta S[\phi_S].
\end{equation}
Including one-loop fluctuations around the saddle point \(\phi_S\) \cite{Coleman:1977,Callan:1977}, this becomes
\begin{equation}
\begin{split}
F_C - F_{\text{symm}}
\simeq{}&
T\, \Delta S[\phi_S]
- T \ln\Bigg[
\left(\frac{\Delta S[\phi_S]}{2\pi}\right)^{3/2} \\
&\times
\left|
\frac{\det'[-\nabla^2+V_3''(\phi_S)]}
{\det[-\nabla^2+V_3''(\phi_0)]}
\right|^{-1/2}
\Bigg]
+ \mathcal{O}(\hbar).
\end{split}
\label{eq:fluctuation}
\end{equation}
Here \(V_3''\) is the second derivative of the 3D effective potential. Thus, the probability ratio contains both the classical action and the one-loop fluctuation correction, without a separate determinant calculation. Numerically, the statistical factor \(A_{\mathrm{stat}}\) is extracted from the difference between the measured suppression exponent \(S\) and the classical action difference \(\Delta S[\phi_S]\) \cite{Moore:2001,Moore:2001b}.

\section{Review of multicanonical sampling}
\label{app:muca}

\subsection{Multicanonical Monte Carlo Method}
\label{app:muca_basic}

Let the Hamiltonian of the system be $H$, and choose a suitable order parameter $\mathcal{O}$. In the canonical ensemble, the probability density of the order parameter $\mathcal{O}$ is
\[
P_{\text{can}}(\mathcal{O}) \propto \int \mathcal{D}\varphi\, e^{-H[\varphi]}\, \delta(\mathcal{O}[\varphi]-\mathcal{O}).
\]
In phase transition or barrier regions, it is often exponentially suppressed, and direct sampling can hardly cover these regions. The multicanonical Monte Carlo method introduces a weight function $W(\mathcal{O})$ that depends only on the order parameter, equivalently defining the multicanonical potential,
\[
V(\mathcal{O}) = -\ln W(\mathcal{O}) + \text{const}.
\]
The sampling weight is modified to
\[
p \propto e^{-H} W(\mathcal{O}) = e^{-H - V(\mathcal{O})},
\]
so the marginal distribution of the order parameter becomes
\[
P_{\text{muc}}(\mathcal{O}) \propto P_{\text{can}}(\mathcal{O})\, e^{-V(\mathcal{O})}.
\]
By iteratively adjusting the weight or the multicanonical potential, one ensures that within the target interval $P_{\text{muc}}(\mathcal{O}) \approx \text{const}$. The optimal form is
\[
V(\mathcal{O}) = \ln P_{\text{can}}(\mathcal{O}) + \text{const},
\]
or equivalently
\[
W(\mathcal{O}) \propto \frac{1}{P_{\text{can}}(\mathcal{O})}.
\]
In this way, the system is forced to perform an approximately uniform random walk within the required range of $\mathcal{O}$, thereby obtaining statistical samples in the exponentially suppressed region. Finally, by reweighting the multicanonical samples with weight $e^{V(\mathcal{O})}$, one recovers the physical probability distribution of the canonical ensemble:
\[
P_{\text{can}}(\mathcal{O}) \propto P_{\text{muc}}(\mathcal{O})\, e^{V(\mathcal{O})}.
\]

\subsection{Effective Action and Multicanonical Potential}
\label{app:muca_effective_action}

Let the lattice field be $\phi$, the canonical action be $S[\phi]$, and the coupling constant be $\beta$. Choose a globally differentiable order parameter $\mathcal{O}[\phi]$. The goal is to produce a flat marginal distribution of the order parameter on the interval $[\mathcal{O}_{\min}, \mathcal{O}_{\max}]$.

In the canonical ensemble, the probability density of the order parameter $\mathcal{O}$ is
\[
P_{\text{can}}(\mathcal{O}) \propto \int \mathcal{D}\phi\, e^{-\beta S[\phi]}\, \delta(\mathcal{O}[\phi]-\mathcal{O}).
\]
Introducing the multicanonical potential $V(\mathcal{O})$, the effective action is taken as
\[
\hat{S}[\phi] = \beta S[\phi] + V(\mathcal{O}[\phi]).
\]
Under this effective action, the sampling weight is $e^{-\hat{S}[\phi]}$, and the corresponding marginal distribution of the order parameter is
\[
P_{\text{MUCA}}(\mathcal{O}) \propto P_{\text{can}}(\mathcal{O})\, e^{-V(\mathcal{O})}.
\]
If it is required to be approximately constant in $[\mathcal{O}_{\min}, \mathcal{O}_{\max}]$, then the optimal multicanonical potential satisfies
\[
V(\mathcal{O}) = \ln P_{\text{can}}(\mathcal{O}) + \text{const}.
\]
In numerical implementations, $V(\mathcal{O})$ is unknown and must be determined through an iterative bootstrap process. Its parameterization can be chosen as piecewise linear or polynomial interpolation, written as
\[
V(\mathcal{O}) = \hat{\beta}(\mathcal{O})\, \mathcal{O} + \hat{\alpha}(\mathcal{O}),
\]
where $\hat{\beta}(\mathcal{O})$ and $\hat{\alpha}(\mathcal{O})$ are undetermined piecewise functions.

\subsection{Weight Iteration and Reweighting}
\label{app:muca_iteration}

The specific steps of weight iteration are as follows:

\begin{enumerate}
\item[(a)] Initial guess: First perform a short canonical HMC simulation to obtain the order-parameter histogram $H_0(\mathcal{O})$. Let the initial weight be $W^{(0)}(\mathcal{O}) \propto 1/H_0(\mathcal{O})$; the corresponding multicanonical potential is
    \[
    V^{(0)}(\mathcal{O}) = -\ln W^{(0)}(\mathcal{O}) + \text{const} = \ln H_0(\mathcal{O}) + \text{const}.
    \]
\item[(b)] Iterative update: Run MCHMC under the current $V^{(i)}(\mathcal{O})$ to obtain a new order-parameter histogram $H^{(i+1)}(\mathcal{O})$. Update the weight
    \[
    W^{(i+1)}(\mathcal{O}) \propto W^{(i)}(\mathcal{O}) \times H^{(i+1)}(\mathcal{O}),
    \]
    or equivalently update the multicanonical potential
    \[
    V^{(i+1)}(\mathcal{O}) = V^{(i)}(\mathcal{O}) - \ln H^{(i+1)}(\mathcal{O}).
    \]
    Repeat this process until $H(\mathcal{O})$ is sufficiently flat in the target interval, e.g., when the ratio of the maximum to the minimum is less than $2$.
\item[(c)] Production run: Fix the converged $V(\mathcal{O})$ and perform large-scale sampling. Finally, recover the canonical ensemble expectation values by reweighting:
    \[
    \langle O \rangle_\beta = \frac{\sum_i O_i\, e^{V(\mathcal{O}_i)}}{\sum_i e^{V(\mathcal{O}_i)}},
    \]
where \(O\) is the physical observation.
\end{enumerate}

\subsection{Multicanonical Hamiltonian Monte Carlo update}
\label{app:mchmc}

HMC uses Hamiltonian dynamics to evolve configurations, effectively reducing autocorrelation times, and has molecular-dynamics trajectories that parallelize well, making it more suitable for large-scale lattice simulations. By specializing the multicanonical framework of Appendix~\ref{app:muca} to the case where an arbitrary differentiable order parameter $\mathcal{O}[\phi]$ is used as the driving variable, one can obtain a multicanonical Hamiltonian Monte Carlo method suitable for lattice field theory and statistical mechanics systems.

Adding the multicanonical effective action \(\hat S[\phi]=\beta S[\phi]+V(\mathcal{O}[\phi])\) to the HMC framework yields an order-parameter-driven multicanonical Hamiltonian Monte Carlo method. Introduce a momentum $p$ conjugate to the field variable, and define the guiding Hamiltonian,
\[
H_{\text{MD}} = \frac{1}{2}\sum p^2 + \hat{S}[\phi]
= \frac{1}{2}\sum p^2 + \beta S[\phi] + V(\mathcal{O}[\phi]).
\]
The accept/reject step uses the same Hamiltonian $H_{\text{acc}} = H_{\text{MD}}$. Differentiating with respect to the field variable gives the force
\[
F(x) = -\frac{\partial \hat{S}}{\partial \phi(x)}
= -\beta \frac{\partial S}{\partial \phi(x)}
- \left.\frac{dV}{d\mathcal{O}}\right|_{\mathcal{O}[\phi]} \frac{\partial \mathcal{O}}{\partial \phi(x)}.
\]
The first term $-\beta\, \partial S/\partial \phi(x)$ is the standard physical force; the second term is the order-parameter biasing force, whose strength is controlled by the global scalar $dV/d\mathcal{O}$, and whose direction is given by $\partial \mathcal{O}/\partial \phi(x)$. For a large class of local or semi-local order parameters, e.g. $\mathcal{O} = \frac{1}{\mathcal{V}} \sum_{\vec{x}} (\phi(\vec{x}) - A)|\phi(\vec{x}) - A|$, one has
\[
\frac{\partial \mathcal{O}}{\partial \phi(x)} = \frac{2}{\mathcal{V}}|\phi(x)-A|.
\]
Therefore, the additional computational overhead of the biasing force is very small. Even for more complex order parameters, as long as their functional derivatives can be computed efficiently, the method remains practical.

The molecular dynamics integration uses a reversible, symplectic leapfrog scheme. Let the step size be $\Delta t$, and the total number of steps be $N_{\text{MD}}$. Each step requires: computing the global order parameter $\mathcal{O}$ from the current configuration; obtaining $dV/d\mathcal{O}$ by table lookup or interpolation; computing the total force $F(x)$; and updating the momenta and fields. A typical update sequence is
\[
p(t+\Delta t/2) = p(t-\Delta t/2) - \Delta t\, F(\phi(t)),
\]
\[
\phi(t+\Delta t) = \phi(t) + \Delta t\, p(t+\Delta t/2),
\]
and half-step corrections are added at the beginning and end of the trajectory to ensure time reversibility.
After the trajectory ends, compute the change in the Hamiltonian,
\[
\Delta H = \hat{S}[\phi'] + \frac{1}{2}\sum p'^2 - \left(\hat{S}[\phi] + \frac{1}{2}\sum p^2\right),
\]
and accept the new configuration with probability
\[
P_{\text{acc}} = \min\left(1, e^{-\Delta H}\right).
\]
This guarantees that the sampling distribution is exactly $e^{-\hat{S}}$, thereby obtaining samples with a flat order parameter. Since the MCHMC sampling is driven by the multicanonical potential, it has good parallel scalability and high ergodic efficiency.

\section{Lattice-continuum matching and continuum extrapolation}
\label{app:lattice_matching}

This appendix collects the lattice-continuum matching and the derivation of the final lattice action used in the main text.

\subsection{High-temperature effective potential and dimensional reduction}
\label{app:dim_red}

We start from the four-dimensional high-temperature effective potential for the electroweak phase transition,
\begin{equation}
V_4(\phi_4)= \frac12 m_4^2(T)\phi_4^2+\frac16 g_4(T)\phi_4^3+\frac{1}{24}\lambda_4\phi_4^4,
\label{eq:potential}
\end{equation}
with
\begin{equation}
m_4^2(T)=\gamma(T^2-T_0^2),\qquad
g_4(T)=-2\alpha T,\qquad
\lambda_4=6\lambda ,
\end{equation}
where \(\phi_4\) is the scalar field, and \(\gamma\), \(T_0\), \(\alpha\), and \(\lambda\) are model dimensionless parameters. The potential is obtained from a high-temperature expansion and is valid when \(T\) is large compared with the field-dependent mass scales.

At high temperature, the equilibrium thermodynamics is dominated by the zero Matsubara mode. The four-dimensional Euclidean action is
\begin{equation}
S_4[\phi_4] = \int_0^\beta d\tau \int d^3x \left[ \frac12 (\partial_\tau \phi_4)^2 + \frac12 (\partial_i \phi_4)^2 + V_4(\phi_4) \right],
\end{equation}
with \(\beta = 1/T\). Expanding \(\phi_4\) in Matsubara modes, we get
\begin{equation}
\phi_4(\tau,\mathbf{x}) = \phi_0(\mathbf{x}) + \sum_{n\neq 0} \phi_n(\mathbf{x}) e^{i\omega_n \tau}, \quad \omega_n = 2\pi n T.
\end{equation}
With only the zero mode \(\phi_0(\mathbf{x})\) kept, the time derivative vanishes and the time integral gives \(\beta = 1/T\), so
\begin{equation}
S_4[\phi_4] \approx \frac{1}{T} \int d^3x \left[ \frac12 (\partial_i \phi_0)^2 + V_4(\phi_0) \right].
\end{equation}
Defining the three-dimensional field
\begin{equation}
\phi_3 = \frac{\phi_0}{\sqrt{T}},
\end{equation}
and the three-dimensional effective potential
\begin{equation}
V_3(\phi_3) = \frac{1}{T} V_4(\sqrt{T}\phi_3, T),
\end{equation}
we obtain the three-dimensional action
\begin{equation}
S_3[\phi_3] = \int d^3x \left[ \frac12 (\partial_i \phi_3)^2 + V_3(\phi_3) \right],
\end{equation}
with \(S_4 \approx S_3\) and the potential \eqref{eq:potential}:
\begin{equation}
\begin{split}
V_3(\phi_3)  
&= \frac12 m_{3,phs}^2 \phi_3^2 + \frac16 g_{3,phs} \phi_3^3 + \frac{1}{24} \lambda_{3,phs} \phi_3^4 \\
&= \frac12 \gamma (T^2 - T_0^2)\phi_3^2 - \frac13 \alpha T^{3/2} \phi_3^3 + \frac14 \lambda T \phi_3^4.
\label{eq:V3_app}
    \end{split}
\end{equation}
This leading-order zero-mode reduction is self-consistent for the class of high-temperature potentials used here, which already incorporate the relevant thermal corrections. Therefore, no further \(T\)-dependent matching is required beyond the zero-mode approximation\cite{Enqvist:1992}.

To make the lattice simulation dimensionless, we introduce a scale $M=150$ GeV and rescale the field and couplings as
\begin{equation}
\begin{split}
&\qquad T_3^0 = T_0 M^{-1}, 
\qquad T_3 = T M^{-1},\\
&\quad \phi=M^{-1/2}\phi_3, 
\quad \lambda_3=M^{-1}\lambda_{3,phs},\\
&\quad m^2_3=M^{-2}m_{3,phs}^2, 
\quad g_3=M^{-3/2}g_{3,phs} .
\label{eq:rescale}
\end{split}
\end{equation}
This rescaling leaves the action invariant but brings the field and couplings into dimensionless lattice units. 

\subsection{Field shift and elimination of the cubic term}
\label{app:shift}

The three-dimensional potential~\eqref{eq:V3_app} contains a cubic term. For the real scalar theory, it is convenient to shift the field by a constant,
\begin{equation}
\phi=\psi+c,\qquad c=-\frac{g_3(T)}{\lambda_3(T)}.
\end{equation}
For fixed \(T\), both \(g_3(T)\) and \(\lambda_3(T)\) are constants, so \(c\) is constant. After the shift, the potential becomes
\begin{equation}
V_3(\psi+c)=\sigma'\psi+\frac12 m'^2(T)\psi^2+\frac{1}{24}\lambda_3(T)\psi^4+\text{constant},
\label{eq:shifted}
\end{equation}
with
\begin{equation}
\sigma'=-\frac{m_3^2(T)g_3(T)}{\lambda_3(T)}
+\frac{g_3(T)^3}{3\lambda_3(T)^2},
\end{equation}
\begin{equation}
m'^2(T)=m_3^2(T)-\frac{g_3(T)^2}{2\lambda_3(T)}.
\end{equation}
The quartic coupling \(\lambda_3(T)\) is unchanged.

The reason for this shift is the following. In the shifted basis, the cubic coupling vanishes, \(g_{\rm lat}=0\). Consequently, the coefficients of odd powers of the field are not renormalised, i.e.
\begin{equation}
\delta\sigma_3=0,\qquad \delta g_3=0 .
\end{equation}
This eliminates the mixing between odd and even operators and simplifies the lattice-continuum matching considerably. One can then extract the lattice-continuum relations for the remaining even couplings from Refs.~\cite{Moore:2001b,ArnoldMoore,Sun}. Although \(m'^2(T)\) and \(\sigma'\) still depend on \(T\), they are constant in a lattice simulation at fixed \(T\). Below we drop the primes and denote the shifted parameters simply by \(\sigma\), \(m^2\), and \(\lambda\).

\subsection{Matching relations}
\label{app:matching}

The bare lattice parameters are related to the \(\overline{\rm MS}\) parameters by
\begin{equation}
\lambda_{\rm lat}=\lambda+\delta\lambda,\qquad
m^2_{\rm lat}=m^2+\delta m^2,\qquad
\sigma_{\rm lat}=\sigma,
\end{equation}
together with the multiplicative factors \(Z_\phi\) and \(Z_m\). The counterterms \(\delta\lambda\), \(\delta m^2\), \(Z_\phi\), and \(Z_m\) are computed in lattice perturbation theory by matching zero-momentum correlation functions: the two- and four-point functions, the leading momentum dependence of the two-point function, and the insertion of \(\langle\phi^2\rangle\). The couplings in three dimensions have positive mass dimension. Each additional loop order reduces the degree of ultraviolet divergence by one, so the matching can be organised order by order~\cite{Moore:2001b,Laine,LaineRajantie}.

For the improved Laplacian~\eqref{eq:improved_laplacian}, the matching relations through \(O(a^2)\) are~\cite{Moore:2001b,GKW}
\begin{equation}
\begin{split}
\delta\lambda&=
\frac{3\xi\lambda^2a}{2(4\pi)}
+\frac{\lambda^3a^2}{(4\pi)^2}
\left(\frac34\xi^2-3C_1-\frac{C_2}{3}\right), \\
Z_\phi&=1+\frac{C_2\lambda^2a^2}{6(4\pi)^2}, \\
Z_m&=1+\frac{\xi\lambda a}{2(4\pi)}
+\frac{\lambda^2a^2}{(4\pi)^2}
\left(\frac{\xi^2}{4}-\frac{C_1}{2}-\frac{C_2}{6}\right), \\
\delta m^2&=
-\frac{\Sigma\lambda}{2(4\pi)a}
+\frac{\lambda^2}{6(4\pi)^2}
\left(\ln\frac{6}{a\mu}+C_3-3\Sigma\xi\right).
\label{eq:continuum_app}
\end{split}
\end{equation}
Here \(\mu\) is the \(\overline{\rm MS}\) renormalisation scale. The numerical constants for the improved Laplacian are~\cite{Moore:2001b}
\begin{equation}
\begin{split}
\Sigma&=2.75238391130752, \\
\xi&=-0.083647053040968, \\
C_1&=0.0550612, \\
C_2&=0.0334416, \\
C_3&=-0.86147916.
\end{split}
\end{equation}
These relations are the $N=1$ Arnold--Moore results and are accurate up to \(O(a^3)\)~\cite{ArnoldMoore,Sun}, except for the mass parameter, which still has an \(O(a)\) residual correction. The operator \(\langle\phi^2\rangle\) also requires additive and multiplicative renormalisation~\cite{Moore:2001b}:
\begin{equation}
a\langle\phi^2\rangle_{\rm cont}
=Z_m\langle\phi^2\rangle_{\rm lat}
-\delta\langle\phi^2\rangle.
\label{eq:phi2_matching_app}
\end{equation}

\subsection{Final lattice action, improved Laplacian and dimensionless rescaling}
\label{app:final_action}

After the above matching and shift, the lattice action takes the form
\begin{equation}
\begin{split}
S_{\rm lat}
&=\sum_x\Big[
-\frac12 Z_\phi\,\phi_x(\nabla^2_{\rm lat}\phi)_x
+\sigma_{\rm lat}\phi_x \\
&\qquad+\frac12 Z_\phi Z_m\,m^2_{\rm lat}\phi_x^2
+\frac{1}{4!}Z_\phi^2\lambda_{\rm lat}\phi_x^4
\Big],
\end{split}
\label{eq:lattice_action_app}
\end{equation}
where \(Z_\phi\) and \(Z_m\) are multiplicative renormalisation constants, and \(\sigma_{\rm lat}\) is the coefficient of the linear term after the shift. The counterterms \(\delta\lambda\) and \(\delta m^2\) are included in the definitions of \(\lambda_{\rm lat}\) and \(m^2_{\rm lat}\) through the matching relations in Appendix~\ref{app:matching}. These counterterms are necessary to cancel lattice discretisation errors and to ensure that the continuum limit reproduces the correct \(\overline{\rm MS}\) physics. 
For the kinetic term, we use the \(O(a^4)\)-improved Laplacian
\begin{equation}
\begin{split}
(\nabla^2_{\rm lat}\phi)_x
&=-\frac{15}{2}\phi_x
+\frac{4}{3}\sum_i(\phi_{x+i}+\phi_{x-i}) \\
&\qquad-\frac{1}{12}\sum_i(\phi_{x+2i}+\phi_{x-2i}),
\end{split}
\label{eq:improved_laplacian}
\end{equation}
whose momentum-space form is
\begin{equation}
\hat k^2=\sum_i\left(\frac52-\frac83\cos k_i+\frac16\cos 2k_i\right).
\label{eq:ksq}
\end{equation}
This improved Laplacian removes the leading \(O(a^2)\) kinetic errors and is also required if the matching is carried to two loops. The lattice-continuum matching relations and the explicit expressions for \(Z_\phi\), \(Z_m\), \(\delta\lambda\), and \(\delta m^2\) are collected in Appendix~\ref{app:matching}. 

\subsection{Overcooling and lattice implementation}
\label{app:overcooling}

The bare lattice mass parameter is related to the continuum $\overline{\rm MS}$ mass by
\begin{equation}
m^2_{\rm lat}=m^2+\delta m^2(a),
\end{equation}
where $\delta m^2(a)$ is computed in lattice perturbation theory (see Eq.~\ref{eq:continuum_app}), which is independent of $m^2$ and depends only on $\lambda_3$ and $a$~\cite{Moore:2001b}. Consequently, although the matching is performed through $O(a^3)$, the mass parameter still receives a residual correction starting at $O(a)$, of the form
\begin{equation}
c_3\lambda_3^3 a+c_4\lambda_3^4 a^2,
\end{equation}
with $c_3,c_4$ pure numbers. These mass-independent lattice artifacts would enter physical observables and make the leading continuum extrapolation $O(a)$ or $O(a^2)$ instead of the desired $O(a^3)$. However, since they are independent of the mass, they cancel exactly in the lattice difference
\begin{equation}
\begin{split}
\Delta m'^2&\equiv m'^2(T)-m_c'^2\\
&=m_{\rm lat}'^2(a)-m_{c,\rm lat}'^2(a),
\label{eq:overcooling_app}
\end{split}
\end{equation}
which is also known as ``overcooling'', where \(m_c'\) is the critical shifted mass.

In the shifted potential~\eqref{eq:shifted}, where the cubic term is eliminated, we have
\begin{equation}
\sigma'=-\frac{m_3^2 g_3}{\lambda_3}+\frac{g_3^3}{3\lambda_3^2},\qquad
m'^2(T)=m_3^2(T)-\frac{g_3(T)^2}{2\lambda_3(T)}.
\end{equation}
The critical temperature is determined by the exact $Z_2$-protected condition
\begin{equation}
\sigma'(T_c)=0,
\label{eq:Tc_condition_app}
\end{equation}
which is not corrected at any order in the loop expansion~\cite{Gould:2021}. This condition is equivalent to
\begin{equation}
m_3^2(T_c)=\frac{g_3(T_c)^2}{3\lambda_3(T_c)}.
\end{equation}
Using the temperature dependence
\begin{equation}
m_3^2(T)=\gamma(T^2-T_0^2),\quad
g_3(T)=-2\alpha T^{3/2},\quad
\lambda_3(T)=6\lambda T,
\end{equation}
Eq.~\eqref{eq:Tc_condition_app} gives
\begin{equation}
T_c^2=\frac{\gamma T_0^2}{\gamma-\dfrac{4\alpha^2}{18\lambda}}.
\end{equation}
The critical shifted mass is then
\begin{equation}
m_c'^2=m_{3,c}^2-\frac{g_3(T_c)^2}{2\lambda_3(T_c)}
=-\frac{g_3(T_c)^2}{6\lambda_3(T_c)}.
\end{equation}
The overcooling is defined as
\begin{equation}
\Delta m'^2\equiv m'^2(T)-m_c'^2,
\end{equation}
see Eq.~\eqref{eq:overcooling_app}.

In the lattice simulation, for each lattice spacing \(a\) we compute the critical bare mass
\begin{equation}
m_{c,\rm lat}'^2(a)=m_c'^2+\delta m'^2(a),
\end{equation}
and set the simulation bare mass to
\begin{equation}
m_{\rm lat}'^2(a)=m_{c,\rm lat}'^2(a)+\Delta m'^2.
\end{equation}
Thus the mass-independent lattice corrections cancel exactly, and the physical overcooling is independent of \(a\). This reduces the leading lattice-spacing dependence of physical observables to \(O(a^3)\), making the continuum extrapolation faster and more reliable.

For the parameters used in this work, we obtain
\begin{equation}
T_c\simeq 1.78294,\quad
m_c'^2\simeq -1.48549,\quad
m'^2(T)\simeq -1.26313,
\end{equation}
and therefore
\begin{equation}
\Delta m'^2\simeq 0.22236.
\end{equation}
In practice, for each \(a\) we set \(m_{\rm lat}'^2(a)=m_{c,\rm lat}'^2(a)+0.22236\).

\end{document}